\documentclass{ws-ijmpd}
\usepackage{color}

\usepackage[colorlinks,%
            citecolor=red,linkcolor=red,hypertex, %
            breaklinks=true]{hyperref}
\begin{document}

\markboth{Xian-Feng Zhao}
{The properties of the massive neutron star PSR J0348+0432}

%
\catchline{}{}{}{}{}
%

\title{The properties of the massive neutron star PSR J0348+0432 }

\author{Xian-Feng Zhao$^{1,2}$
\footnote{School of Sciences, Southwest Petroleum University, Chengdu, 610500, China}}

\address{$^{1}$ School of Sciences, Southwest Petroleum University, Chengdu, 610500, China\\
$^{2}$ School of Electronic and Electrical Engineering, Chuzhou University, Chuzhou, 239000, China\\
zhaopioneer.student@sina.com}



\maketitle

\begin{history}
\received{February 4, 2015}
\revised{\today}
\comby{Managing Editor}
\end{history}

\begin{abstract}
The properties of the massive neutron star PSR J0348+0432 is calculated in the framework of the relativistic mean field theory by choosing the suitable hyperon coupling constants. It is found that the central energy density $\epsilon_{c}$ and the central pressure $p_{c}$ of the massive neutron star PSR J0348+0432 respectively are 1.5 times larger and 3.6 times larger than those of the canonical mass neutron star. It is also found that in the neutron star PSR J0348+0432 there are five kinds of baryons appearing: n, p, $\Lambda$, $\Xi^{-}$ and $\Xi^{0}$ but in the canonical mass neutron star there are only three kinds of particles appearing: n, p and $\Lambda$. In our models, the positive well depth $U_{\Sigma}^{(N)}$ will restrict the production of the hyperons $\Sigma^{-}$, $\Sigma^{0}$ and $\Sigma^{+}$ and therefore either in the neutron star PSR J0348+0432 or in the canonical mass neutron star the hyperons $\Sigma^{-}$, $\Sigma^{0}$ and $\Sigma^{+}$ all do not appear. In addition, our results also show that the radius $R$ of the massive neutron star PSR J0348+0432 is less than that of the canonical mass neutron star while the gravitational redshift of the former is larger than that of the latter.
\end{abstract}


{\color{blue}\emph{PACS numbers:}
21.65.+f, 24.10.Jv,26.60.+c,21.30.Fe}  

\section{Introduction}
Recent years, several massive neutron stars have been discovered successively. In 2010 the neutron star PSR J1614-2230 with the mass of 1.97$\pm$0.04 M$_{\odot}$ was observed~\cite{Demorest10}. Through a causality equation $R\geq8.3(\frac{M}{1.97M_{\odot}})$ km, in conjunction with the recent spectroscopic measurements of neutron-star radii in X-ray binaries that place them at values 12 km~\cite{{Ozelprd10},{Steiner10}}, it would lead to the very narrow range 8.3 km$\leq R \leq$ 12 km for all neutron stars~\cite{Ozel10}. Hereafter, the massive neutron stars have been widely studied with different models of dense matter, such as massive hadronic neutron stars~\cite{{Zhao12},{Bednarek12},{Bednarek13},{Bejger13},{Jiang12},{Miyatsu13},{WeissenbornNPA12},{WeissenbornPRC12}}, quark-hadron hybrid stars~\cite{{Klahn13},{Last11},{Masuda12}}, hadronic stars within other approximations such as mean-field relativistic or quark-meson coupling model~\cite{{Massot12},{Whittenbury12},{Katayama12}} and so on.

It is well known that the gravitational redshift of a neutron star is closely connected to the value of $M/R$, with $M$ being the mass and $R$ the corresponding radius. Supposing that the gravitational redshift is known, if the mass $M$ or the radius $R$ is known the other quantity would be determined~\cite{Glen97}. To the contrary, if the mass $M$ and the radius $R$ are all known the gravitational redshift also can be calculated.

The mass and the radius of a neutron star are interconnected by Oppenheimer-Volkoff (O-V) equation. Since the coupling constants would influence the properties of the neutron star matter, once the mass is known it will provide constraints on the properties of the neutron star matter and this would constrain the coupling constants, which would influence the radius. So, each newly observed neutron star mass would provide a new constraint on the radius and further on the gravitational redshift.

In 2013, massive neutron star PSR J0348+0432 with the mass of $2.01\pm0.04$ M$_{\odot}$ was obsearved by Antoniadis et al~\cite{Anton13}. Its massive mass must undoubtedly provide constraints on the radius and the gravitational redshift. The past studies have found the gravitational redshift of the neutron stars is in the range of 0.25 $\sim$ 0.35~\cite{Liang86}. But for the massive neutron star PSR J0348+0432, its gravitational redshift is still not known.

The observation data of the gravitational redshift of the massive neutron star PSR J0348+0432 is lacking and so to theoretically study it is necessary. For this purpose, its mass together with its radius must first be obtained theoretically.

In order to obtain a more lager neutron star mass, the equation of state (EoS) should be enough harder. The calculations showed the mass of neutron stars is in the range of $1.5 \sim 1.97$ M$_{\odot}$ if the hyperon degree of freedom is considered but can gain 2.36 M$_{\odot}$ as only the nucleons being considered~\cite{Glen85,Glen91,Zhao12}. Obviously, the emergence of the hyperons will reduce the neutron star mass.

Neutron stars are compact objects, within which hyperons will appear. So considering the hyperon degree of freedom in neutron stars should be reasonable~\cite{Glen85}. To calculate the mass of a neutron star by the relativistic mean field (RMF) theory, we must choose the nucleon coupling constants and the hyperon coupling constants at first.

The calculations showed that the nucleon coupling constants GL85 can better describe the properties of the neutron star matter~\cite{Glen97}. For the hyperon coupling constants, there are many selection method. For example, we can select the hyperon coupling constants of mesons $\rho, \omega$ by SU(6) symmetry and those of meson $\sigma$ by fitting the $\Lambda, \Sigma$ and $\Xi$ well depth in nuclear matter~\cite{Zhao12}.

In this paper, the properties of the massive neutron star PSR J0348+0432 is studied in the framework of the RMF theory considering the baryon octet.

\section{The RMF theory and the mass of a neutron star}
The Lagrangian density of hadron matter reads as follows~\cite{Glen97}
\begin{eqnarray}
\mathcal{L}&=&
\sum_{B}\overline{\Psi}_{B}(i\gamma_{\mu}\partial^{\mu}-{m}_{B}+g_{\sigma B}\sigma-g_{\omega B}\gamma_{\mu}\omega^{\mu}
\nonumber\\
&&-\frac{1}{2}g_{\rho B}\gamma_{\mu}\tau\cdot\rho^{\mu})\Psi_{B}+\frac{1}{2}\left(\partial_{\mu}\sigma\partial^{\mu}\sigma-m_{\sigma}^{2}\sigma^{2}\right)
\nonumber\\
&&-\frac{1}{4}\omega_{\mu \nu}\omega^{\mu \nu}+\frac{1}{2}m_{\omega}^{2}\omega_{\mu}\omega^{\mu}-\frac{1}{4}\rho_{\mu \nu}\cdot\rho^{\mu \nu}
\nonumber\\
&&+\frac{1}{2}m_{\rho}^{2}\rho_{\mu}\cdot\rho^\mu-\frac{1}{3}g_{2}\sigma^{3}-\frac{1}{4}g_{3}\sigma^{4}
\nonumber\\
&&+\sum_{\lambda=e,\mu}\overline{\Psi}_{\lambda}\left(i\gamma_{\mu}\partial^{\mu}
-m_{\lambda}\right)\Psi_{\lambda}
.\
\end{eqnarray}
This Lagrangian is the non-linear Walecka-Boguta-Bodmer type. Here, $\Psi_{B}$ is the Dirac spinor of the baryon B, the corresponding mass is $m_{B}$. $\sigma, \omega$ and $\rho$ are the field operators of the $\sigma, \omega$ and $\rho$ mesons, respectively. $m_{\sigma}$, $m_{\omega}$ and $m_{\rho}$ are the masses
of these mesons and $m_{\lambda}$ expresses the leptonic mass. $g_{\sigma_{B}}$, $g_{\omega_{B}}$ and $g_{\rho_{B}}$ are, respectively, the coupling constants of the $\sigma$, $\omega$ and $\rho$ mesons with the baryon B. $\frac{1}{3}g_{2}\sigma^{3}+\frac{1}{4}g_{3}\sigma^{4}$ expresses the self-interactive energy, in which $g_{2}$ and $g_{3}$ are the self-interaction parameters of $\sigma$-meson. And the last term expresses the Lagrangian of both electron and muon.

Then the RMF approach is used. The condition of $\beta$ equilibrium in neutron star matter demands the chemical equilibrium
\begin{eqnarray}
\mu_{i}=b_{i}\mu_{n}-q_{i}\mu_{e},
\end{eqnarray}
where $b_{i}$ is the baryon number of a species $i$ and $q_{i}$ is its electric charge number.

We choose the nucleon coupling constant GL85 set in this work~\cite{Glen85} : the saturation density $\rho_{0}$=0.145 fm$^{-3}$, binding energy B/A=15.95 MeV, a compression modulus $K=285$ MeV, charge symmetry coefficient $a_{sym}$=36.8 MeV and the effective mass $m^{*}/m$=0.77.

For the hyperon coupling constant, we define the ratios: $x_{\sigma h}=\frac{g_{\sigma h}}{g_{\sigma}}=x_{\sigma}, x_{\omega h}=\frac{g_{\omega h}}{g_{\omega}}=x_{\omega}, x_{\rho h}=\frac{g_{\rho h}}{g_{\rho}},$ with $h$ denoting hyperons $\Lambda, \Sigma$ and $\Xi$.

We choose $x_{\rho \Lambda}=0, x_{\rho \Sigma}=2, x_{\rho \Xi}=1$ by SU(6) symmetry~\cite{Schaff96}. The experiments show $U_{\Lambda}^{(N)}=-30$ MeV~\cite{Batt97}, $ U_{\Sigma}^{(N)}=10\sim40$ MeV~\cite{Kohno06,Harada05,Harada06,Fried07} and $U_{\Xi}^{(N)}=-28$ MeV~\cite{Schaff00}. Therefore, in this work we choose $U_{\Lambda}^{(N)}=-30$ MeV, $ U_{\Sigma}^{(N)}$=40 MeV and $U_{\Xi}^{(N)}=-28$ MeV, respectively.

The calculations show the ratio of hyperon coupling constant to nucleon coupling constant is in the range of $\sim$ 1/3 to 1~\cite{Glen91}. So, we choose $x_{\sigma}$=0.4, 0.5, 0.6, 0.7, 0.8, 0.9,1.0 and for each $x_{\sigma}$ we first choose $x_{\omega}$=0.4, 0.5, 0.6, 0.7, 0.8, 0.9,1.0, respectively. Then, considering the restriction of the hyperon well depth~\cite{Glen97}

\begin{eqnarray}
U_{h}^{(N)}=m_{n}\left(\frac{m_{n}^{*}}{m_{n}}-1\right)x_{\sigma}+\left(\frac{g_{\omega}}{m_{\omega}}\right)^{2}\rho_{0}x_{\omega}
,\
\end{eqnarray}
the hyperon coupling constants $x_{\omega}$ will be slightly adjusted. Thus parameters we choose are listed in Table~\ref{tab1}.

\begin{table}[!htbp]
\tbl{The hyperon coupling constants fitted to the experimental data of the well depth, which are $U_{\Lambda}^{N}=-30$ MeV, $U_{\Sigma}^{N}=+40$ MeV and $U_{\Xi}^{N}=-28$ MeV, respectively.}
{\begin{tabular}{@{}cccccccc@{}} \toprule
$x_{\sigma \Lambda}$ &$x_{\omega \Lambda}$&$x_{\sigma \Sigma}$  &$x_{\omega \Sigma}$ &$x_{\sigma \Xi}$     &$x_{\omega \Xi}$    \\
\hline
0.4               &0.3679                  &0.4               &0.8250              &0.4               &0.3811              \\
0.5               &0.5090                  &\underline{0.5}   &\underline{0.9660}  &0.5               &0.5221                \\
0.6               &0.6500                  &                  &                    &0.6               &0.6630              \\
0.7               &0.7909                  &                  &                    &0.7               &0.8040              \\
0.8               &0.9319                  &                  &                    &0.8               &0.9450              \\
\hline
\end{tabular} \label{tab1}}
\end{table}

We use the O-V equation to obtain the mass and the radius of neutron stars
\begin{eqnarray}
\frac{\mathrm dp}{\mathrm dr}&=&-\frac{\left(p+\varepsilon\right)\left(M+4\pi r^{3}p\right)}{r \left(r-2M \right)}
,\\\
M&=&4\pi\int_{0}^{r}\varepsilon r^{2}\mathrm dr
.\
\end{eqnarray}

\begin{figure}[!htp]
\begin{center}
\includegraphics[width=3.5in]{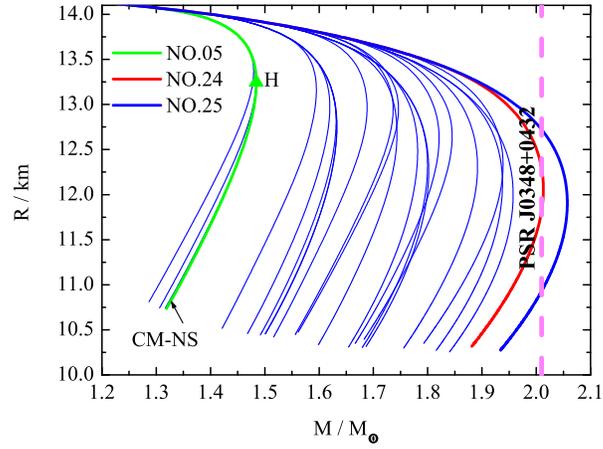}
\caption{The radius as a function of the mass of the neutron star.}
\label{fig1}
\end{center}
\end{figure}

In the present model, $\Sigma$ hyperons are not produced anyway~\cite{Zhao11}. So in Table~\ref{tab1} we only choose $x_{\sigma \Sigma}=0.4$ and $x_{\omega \Sigma}$=0.825 while $x_{\sigma \Sigma}=0.5$ and $x_{\omega \Sigma}$=0.9660 can be deleted. From Table~\ref{tab1}, we can compose of 25 sets of suitable parameters ( named as NO.01, NO.02, ..., NO.25, respectively ), for which we calculate the mass of the neutron star. The results show only sets of parameters NO.24 ( $x_{\sigma \Lambda}=0.8, x_{\omega \Lambda}$=0.9319; $x_{\sigma \Sigma}=0.4, x_{\omega \Sigma}=0.825$; $x_{\sigma \Xi}=0.7, x_{\omega \Xi}=0.804$ ) and NO.25 ( $x_{\sigma \Lambda}=0.8, x_{\omega \Lambda}$=0.9319; $x_{\sigma \Sigma}=0.4, x_{\omega \Sigma}=0.825$; $x_{\sigma \Xi}=0.8, x_{\omega \Xi}=0.945$ ) can obtain the masses greater than that of the massive neutron star PSR J0348+0432 ( for the former, the maximum mass is $M_{max}$=2.0132 $M_{\odot}$, for the latter, the maximum mass is $M_{max}$=2.0572 $M_{\odot}$) ( see Fig.~\ref{fig1}).

\begin{figure}[!htp]
\begin{center}
\includegraphics[width=3.5in]{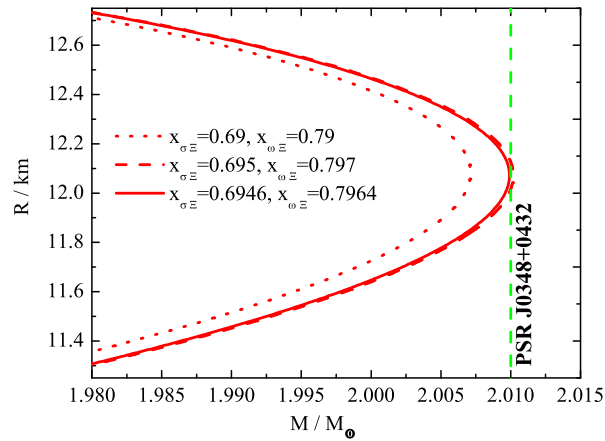}
\caption{The fitting process of the neutron star PSR J0348+0432.}
\label{fig2}
\end{center}
\end{figure}

We see that the neutron star mass obtained by NO.24 is $M_{max}$=2.0132 M$_{\odot}$, which is more close to the mass of the neutron star PSR J0348+0432 ( $M$=2.01 M$_{\odot}$ ). Next, we respectively fine tune the $x_{\sigma \Xi}$ to 0.69, 0.695, 0.6946 and the corresponding $x_{\omega \Xi}$ got by fitting to the well depth are 0.79, 0.797, 0.7964, respectively. With the parameters above, the corresponding maximum mass of the neutron star respectively are $M_{max}$=2.007, 2.0102, 2.01 M$_{\odot}$. Thus, we have obtained one set of parameters which can give the mass of the neutron star PSR J0348+0432: $x_{\sigma \Lambda}=0.8$, $x_{\omega \Lambda}=0.9319$, $x_{\sigma \Sigma}=0.4$, $x_{\omega \Sigma}=0.825$, $x_{\sigma \Xi}=0.6946$, $x_{\omega \Xi}=0.7964$( see Fig.~\ref{fig2}). In the next step, with it we can describe the gravitational redshift of the neutron star PSR J0348+0432.

Besides, because parameter NO.05 ( $x_{\sigma \Lambda}=0.4, x_{\omega \Lambda}$=0.3679; $x_{\sigma \Sigma}=0.4, x_{\omega \Sigma}=0.825$; $x_{\sigma \Xi}=0.8, x_{\omega \Xi}=0.945$ ) gives the maximum mass $M=$1.4843 M$_{\odot}$, which is close to the mass of the canonical mass neutron star ( named as CM-NS ), we choose it as a comparison. Although there is no freedom to change the coupling constants of the hypernuclear Lagrangian to describe massive and canonical stars, in the two cases the neucleon copling constant GL85 is yet the same. We think that to comparise these two cases still is meaningful.

\section{The properties of the massive neutron star PSR J0348+0432}
The properties of the neutron stars calculated in this work is shown in Table~\ref{tab2}, from which and Fig.~\ref{fig2} we see the radius of the massive neutron star PSR J0348+0432 is $R=12.072$ km but that of the canonical mass neutron star is $R=13.245$ km in our calculations. The former is about 9 \% less than the latter.

\begin{figure}[!htp]
\begin{center}
\includegraphics[width=3.5in]{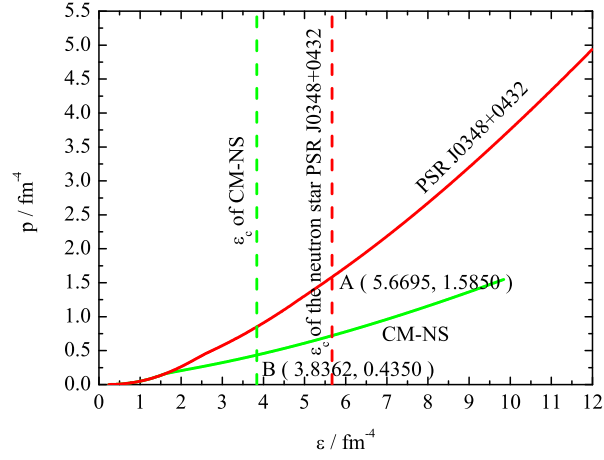}
\caption{The pressure as a function of the energy density. The $\varepsilon_{c}$ denotes the central energy density of the neutron star.}
\label{fig3}
\end{center}
\end{figure}

The mass of the massive neutron star PSR J0348+0432 is $M=$ 2.01 M$_\odot$ while that of the canonical mass neutron star is only $M=$ 1.48 M$_\odot$ in our work, i.e. the former is about 1.36 times larger than the latter. This means the EoS of the massive neutron star PSR J0348+0432 must be more harder than that of the canonical mass neutron star. These cases can be seen in Fig.~\ref{fig3}. In addition, the central energy density of the massive neutron star PSR J0348+0432 is $\epsilon_{c}$=5.6695 fm$^{-4}$ while that of the canonical mass neutron star is only $\epsilon_{c}$=3.8362 fm$^{-4}$, i.e. the former is about 1.5 times larger than the latter. We also can see that the central pressure of the massive neutron star PSR J0348+0432 is $p_{c}$=1.585 fm$^{-4}$ while that of the canonical mass neutron star is only $p_{c}$=0.435 fm$^{-4}$, which means the former is about 3.6 times larger than the latter. So harder EoS of the neutron star PSR J0348+0432 necessarily provide more lager mass.

\begin{table}[!htbp]
\tbl{The mass $M$, the radius $R$, the central energy density $\varepsilon_{c}$, the central pressure $p_{c}$, the central baryon density $\rho_{c}$ and the gravitational redshift $z$ of the neutron stars calculated in this work.}
{\begin{tabular}{@{}ccccccc@{}} \toprule
neutron star&$M$        &$R$   &$\varepsilon_{c}$&$p_{c}$  &$\rho_{c}$ &$z$        \\
            &M$_{\odot}$&km    &fm$^{-4}$        &fm$^{-4}$&fm$^{-3}$  &      \\
\hline
PSR J0348+0432 &2.01    &12.072&5.6695           &1.585    &0.9153 &0.4026         \\
CM-NS          &1.48    &13.245&3.8362           &0.435    &0.6961 &0.2226          \\
\hline
\end{tabular} \label{tab2}}
\end{table}

\begin{figure}[!htp]
\begin{center}
\includegraphics[width=3.5in]{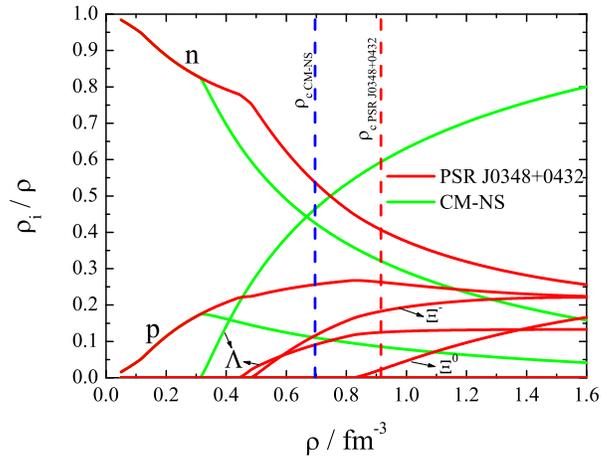}
\caption{The relative particle number density as a function of the baryon density. The $\rho_{c}$ denotes the central baryon density.}
\label{fig4}
\end{center}
\end{figure}

On the other hand, the difference between the EoS of the neutron star PSR J0348+0432 and that of the canonical mass neutron star must indicate the difference of the particles distribution in the star. Fig.~\ref{fig4} gives the relative particle number density as a function of the baryon density.
We see that in the neutron star PSR J0348+0432 there are five kinds of baryons appearing: n, p, $\Lambda$, $\Xi^{-}$ and $\Xi^{0}$ but in the canonical mass neutron star there are only three kinds of particles appearing: n, p and $\Lambda$. Because of the lower density in the canonical mass neutron star the hyperons $\Xi^{-}$ and $\Xi^{0}$ can not produce. In our models, the positive well depth $U_{\Sigma}^{(N)}$ will restrict the production of the hyperons $\Sigma^{-}$, $\Sigma^{0}$ and $\Sigma^{+}$. Therefore, either in the neutron star PSR J0348+0432 or in the canonical mass neutron star the hyperons $\Sigma^{-}$, $\Sigma^{0}$ and $\Sigma^{+}$ all do not appear.

\section{The gravitational redshift of the massive neutron star PSR J0348+0432}

The gravitational redshift of a neutron star is given by~\cite{Glen97}
\begin{eqnarray}
z=(1-\frac{2M}{R})^{-1/2}-1.
\end{eqnarray}

From the above we see the difference between the EoS of the neutron star PSR J0348+0432 and that of the canonical mass neutron star results in the differences of the mass and the radius. Because the gravitational redshift is connected with the ratio of mass to radius $M/R$ ( see Eq.(6)), the difference between the EoS inevitably lead to the difference of the gravitational redshift between them.

\begin{figure}[!htp]
\begin{center}
\includegraphics[width=3.5in]{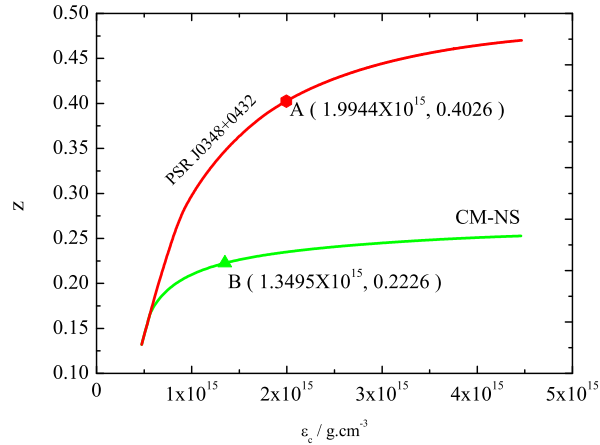}
\caption{The gravitational redshift as a function of the central energy density.}
\label{fig5}
\end{center}
\end{figure}

The gravitational redshift $z$ as a function of the central energy density $\epsilon_{c}$ is given in Fig.~\ref{fig5}. We see the gravitational redshift of the massive neutron star PSR J0348+0432 is $z=$0.4026 while that of the canonical mass neutron star is $z=$0.2226. The former is about 1.8 times larger than the latter. The value range of the gravitational redshift given by Liang is 0.25$\sim$0.35~\cite{Liang86}. Obviously, the gravitational redshift of the massive neutron star PSR J0348+0432 is far greater than that value range. That is because the mass of the neutron star PSR J0348+0432 is more larger than those observed before.

\begin{figure}[!htp]
\begin{center}
\includegraphics[width=3.5in]{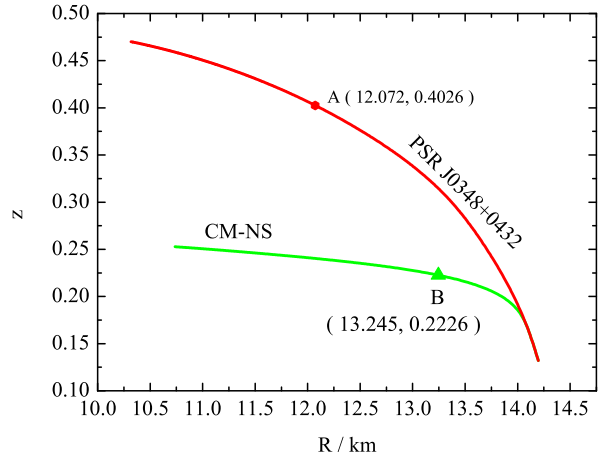}
\caption{The gravitational redshift as a function of the radius.}
\label{fig6}
\end{center}
\end{figure}

\begin{figure}[!htp]
\begin{center}
\includegraphics[width=3.5in]{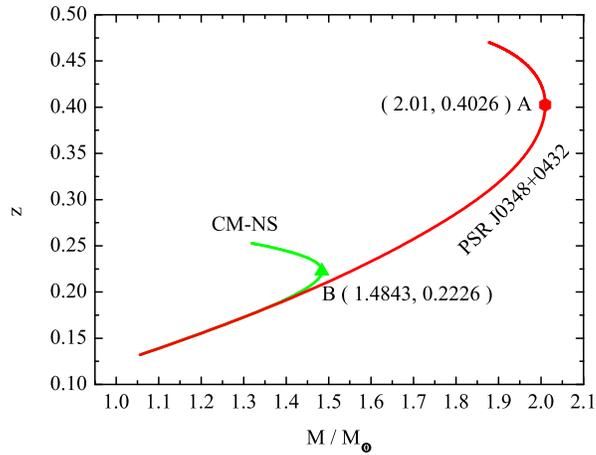}
\caption{The gravitational redshift as a function of the mass.}
\label{fig7}
\end{center}
\end{figure}

Figure~\ref{fig6} and Fig.~\ref{fig7} display the gravitational redshift $z$ as a function of the radius $R$ or the mass $M$, respectively. We see the gravitational redshift $z$ decreases with the radius $R$ increases but increase with the mass $M$ increases. From Eq.(6) we know the gravitational redshift $z$ is positive correlation with the ratio $M/R$. For the neutron star PSR J0348+0432 the ratio is $M/R$=0.1665 while for the canonical neutron star the ratio is $M/R$=0.1121. The former is greater that the latter. So the gravitational redshift of the massive neutron star PSR J0348+0432 is greater than that of the canonical mass neutron star.

In this work, by adjusting the hyperon couplings we obtain one set of parameters, with which we studied the the gravitational redshift of the massive neutron star PSR J0348+0432. But, of course, through the similar steps we also can get a lot of other sets of parameter, which can similarly give the mass of the neutron star PSR J0348+0432 and equally fit to the hyperon well depth. Hence, the parameters used in this work only is one of all the possible ones. But even so, our results to some extent can illustrate the properties especially the gravitational redshift of the massive neutron star PSR J0348+0432.
\section{Summary}
In this paper, the properties of the massive neutron star PSR J0348+0432 is calculated in the framework of the RMF theory by choosing the suitable hyperon coupling constants. It is found that if $U_{\Lambda}^{(N)}=-30$ MeV, $ U_{\Sigma}^{(N)}$=40 MeV, $U_{\Xi}^{(N)}=-28$ MeV, $x_{\rho \Lambda}=0, x_{\rho \Sigma}=2, x_{\rho \Xi}=1$ and the nucleon coupling constant GL85 are chosen we can obtain one set of suitable parameter, which can describe the mass of the massive neutron star PSR J0348+0432.

The results show the central energy density $\epsilon_{c}$ of the massive neutron star PSR J0348+0432 is 1.5 times larger than that of the canonical mass neutron star and the central pressure $p_{c}$ of the massive neutron star PSR J0348+0432 is 3.6 times larger than that of the canonical mass neutron star. We also see that in the neutron star PSR J0348+0432 there are five kinds of baryons appearing: n, p, $\Lambda$, $\Xi^{-}$ and $\Xi^{0}$. But in the canonical mass neutron star, only three kinds of particles ( n, p and $\Lambda$ ) appear and the hyperons $\Xi^{-}$ and $\Xi^{0}$ do not produce. Either in the neutron star PSR J0348+0432 or in the canonical mass neutron star the hyperons $\Sigma^{-}$, $\Sigma^{0}$ and $\Sigma^{+}$ all do not appear.

We also can see the radius and the gravitational redshift of the massive neutron star PSR J0348+0432 respectively are $R$=12.072 km and $z$=0.4026 and those of the canonical mass neutron star severally are $R$=13.245 km and $z$=0.2226. That is to say, the radius of the massive neutron star PSR J0348+0432 is about 9 \% less than that of the canonical mass neutron star and the gravitational redshift of the massive neutron star PSR J0348+0432 is about 1.8 times larger than that of the canonical mass neutron star.

The recent published work on hyperonic dense matter and compact stars show that to assume the massive neutron stars contain hyperon freedom perhaps is reasonable~\cite{{Colucci13},{van Dalen14},{Oertel14},{Raduta14}}. These cases are consistent with our results.

The results about the neutron star PSR J0348+0432 should be also applicable to "massive neutron stars", which are defined as those with masses close to 2 M$_{\odot}$.

\section*{Acknowledgments}
We are thankful to the anonymous referee for many useful comments and suggestions. This work was supported by the Special Funds for Theoretical Physics Research Program of the Natural Science Foundation of China ( Grant No. 11447003 ) and the Scientific Research Foundation of the Higher Education Institutions of Anhui Province, China ( Grant No. KJ2014A182 ).

\end{document}